\documentstyle[12pt,epsf,epsfig]{article}

\begin{document}

\begin{flushright}
{\bf FERMILAB-Pub-00/195-T}\\
\today
\end{flushright}
\vspace{0.5cm}

\begin{center}

\huge
{\bf Neutrino Factories:  Physics}
\end{center}
\bigskip
\vspace{1.0cm}
\large
\begin{center}
Steve Geer

Fermi National Accelerator Laboratory, PO Box 500, Batavia, IL60137, USA
\end{center}
\vspace{0.4cm}
\begin{center}
{\bf Abstract}
\end{center}
\bigskip
\begin{center}
The recent evidence for neutrino oscillations opens a new 
and exciting era in neutrino physics. The next generation 
of accelerator based 
neutrino oscillation experiments are expected to confirm 
the nature of the oscillations, and begin to measure some of the associated 
oscillation parameters. However, these experiments will not be able
to completely determine the mixing matrix, determine the pattern of 
neutrino masses, or search for CP violation in the lepton sector. 
Therefore we are motivated to consider the neutrino beams that will be 
needed beyond the next generation of experiments. 
With this in mind the physics case for a neutrino factory is discussed. 
It is shown that this new type of neutrino source would enable the 
crucial questions to be addressed, and perhaps provide enough information 
to discriminate between Grand Unified Theories, or lead us to an 
alternative theoretical framework. It is possible that measurements at 
a neutrino factory will point the way towards 
an understanding of the origin of quark and lepton flavors.

\end{center}
 
\normalsize
\newpage

\section{Prologue: Intense muon sources}

In recent years there has been much interest in the possibility 
of developing a new generation of very intense muon sources capable 
of producing a millimole of muons per year. This interest is well 
motivated. A very intense muon source producing a bright beam 
that can be rapidly accelerated to high energies would provide a 
new tool for particle physics. At present the beam toolkit 
available for physicists interested in particle interactions 
at the highest energies is limited to beams of charged stable 
particles: electrons, positrons, protons, and antiprotons. 
The development of intense bright $\mu^+$ and $\mu^-$ beams would extend this 
toolkit in a significant way, opening the door for multi-TeV 
muon colliders, lower energy muon colliders (Higgs factories), 
muon--proton colliders, etc. In addition, all of the muons decay 
to produce neutrinos. Hence a new breed of high energy and high 
intensity neutrino beams would become possible. Finally, there is 
the prospect of using the low energy (or stopped) muons to 
study rare processes with orders of magnitude more muons than 
currently available.

In response to the seductive vision of a millimole muon source an 
R\&D collaboration was formed in the US in 1995, initially 
motivated by the desire to design a multi-TeV muon collider, 
and more recently by the desire to design a ``neutrino factory'' 
as a step towards a muon collider. 
The design concepts for a neutrino factory facility are described 
in the accompanying article written by Andrew Sessler~\cite{andy}. 
The motivation for neutrino factories is two-fold. 
First, the neutrino physics that could be pursued at a neutrino 
factory is compelling: the subject of this article. 
Second, a neutrino factory would provide a physics-driven 
project that would facilitate the development of millimole muon 
sources: the enabling technology for so many other goodies, 
including muon colliders.

\section{Why do we need a new neutrino source ?}

Results from the Superkamiokande experiment~\cite{superk} (SuperK) have 
yielded convincing evidence for a deficit of muon-type 
neutrinos ($\nu_\mu$) in the atmospheric neutrino flux. 
This deficit varies with the zenith angle of the incident neutrinos, and 
hence varies with the distance between the source 
and the detector. The natural interpretation of this 
result is that the missing $\nu_\mu$ have oscillated into 
$\nu_X$ as they traversed the distance $L$ between 
their point of production in the atmosphere and the 
detector. The final state flavor $\nu_X$ is currently 
believed to be $\nu_\tau$ since (i) the appropriate region
of parameter space for $\nu_\mu \rightarrow \nu_e$ oscillations 
is already excluded by the CHOOZ experiment~\cite{chooz}, and 
(ii) oscillations into a sterile neutrino $\nu_S$ are excluded 
at the 99\% Confidence Level by other SuperK measurements. 

The SuperK results open a new and exciting era in neutrino physics. 
Neutrino oscillation experiments are no longer searches for a 
phenomenon that may or may not exist. The experimental sensitivity 
required to measure oscillations is now known, and the great thing is 
that $\nu_\mu \rightarrow \nu_X$ oscillations are 
within reach of the next generation of accelerator based experiments.  
Why is this exciting ?  The reason is that, since neutrinos oscillate, 
they must have mass, requiring either the existence of right handed 
neutrinos (Dirac masses) or lepton number violation (Majorana masses), 
or both. Hence, neutrino oscillations cannot be accommodated within the 
Standard Model. The origin 
of neutrino masses must arise from physics beyond the Standard Model. 
Theories that describe physics beyond the Standard Model at Grand Unified 
scales (GUTs) predict 
patterns of oscillation parameters (mixing angles and neutrino masses). 
Comprehensive measurements of neutrino oscillations can therefore 
discriminate between GUTs. 
Note that GUTs also ``predict'' the number of quark and lepton 
generations. Perhaps neutrino oscillation measurements will help 
us understand why there are three families. 
In addition, precision neutrino oscillation 
measurements can determine, or put stringent limits on, CP violation 
in the lepton sector. So it appears that we now have, within reach 
of a new generation of accelerator based experiments, an exciting window 
on physics at the GUT scale, CP violation in the lepton sector, 
the origin of neutrino masses and, perhaps, 
the origin of quark and lepton flavors.

As if this were not motivation enough for detailed neutrino 
oscillation studies, there is more. 
First, there is the long standing solar neutrino problem: a 
deficit of neutrinos from the sun compared to the predictions of 
the Standard Solar Model. This discrepancy might also be due to 
neutrino oscillations, in this case the oscillations 
$\nu_e \rightarrow \nu_x$. 
In the next few years results from the SNO~\cite{sno} and 
KamLAND~\cite{kamland} experiments 
are expected to strengthen the evidence for (or reject) solar 
neutrino oscillations. 
If accelerator based experiments can subsequently 
measure all of the parameters associated with neutrino oscillations 
we may very well resolve the solar neutrino problem. 
Second, there is evidence for $\nu_\mu \rightarrow \nu_e$ and 
$\overline{\nu}_\mu \rightarrow \overline{\nu}_e$ oscillations from 
an accelerator experiment (LSND~\cite{lsnd}) at Los Alamos. 
The problem here is 
that the splittings between the participating neutrino mass eigenstates 
needed to explain the atmospheric neutrino deficit, the solar neutrino 
deficit, and the LSND result, are all different. If all three results 
are due to neutrino oscillations we need three different mass splittings. 
However, we know of only three neutrino flavors, which can 
accommodate at most two mass splittings. There is the shocking possibility 
that there are additional neutrino flavors: sterile neutrinos. This leads 
us to a further motivation for detailed neutrino oscillation studies, 
namely to determine whether light sterile neutrinos exist.

With all of these incentives, we can ask: What neutrino beams will be  
needed in the future to determine all of the oscillation parameters, 
constrain GUT scale theories, learn about CP violation in the lepton 
sector, resolve the solar neutrino problem, and determine whether 
there are light sterile neutrinos?  
In the following sections we will see that we will certainly need higher 
intensity beams than already foreseen. We will also need beams 
propagating through the Earth over baselines of several thousand kilometers, 
and it is 
probably essential, and certainly highly desirable, that we have 
$\nu_e$ and $\overline{\nu}_e$ beams in addition to 
$\nu_\mu$ and $\overline{\nu}_\mu$ beams.

\section{Why neutrino factories ?}

Conventional neutrino beams are produced from a beam of 
charged pions decaying in a long (typically several hundred meters) 
decay channel. If positive (negative) pions are selected, the result is an 
almost pure $\nu_\mu$ ($\overline{\nu}_\mu$) beam 
from $\pi^+ \rightarrow \mu^+ \nu_\mu$   
($\pi^- \rightarrow \mu^- \overline{\nu}_\mu$) decays, 
with a small O(1\%) component of $\nu_e$ from three body kaon decays. 
The $\nu_e$ component is not large enough to be useful for 
$\nu_e \rightarrow \nu_X$ measurements.
Hence, if we want $\nu_e$ and $\overline{\nu}_e$ beams we will need a 
different sort of neutrino source.

An obvious way to try to get $\nu_e$ and $\overline{\nu}_e$ beams 
is to exploit 
the decays $\mu^+ \rightarrow e^+ \nu_e \overline{\nu}_\mu$ and 
$\mu^- \rightarrow e^- \nu_\mu \overline{\nu}_e$. 
To create a neutrino beam with sufficient intensity for a new generation 
of oscillation experiments will require a very intense muon source. 
With a millimole of muons per year we can imagine producing high energy 
beams containing O($10^{20}$) neutrinos and antineutrinos per year. 
However, to achieve this  
a large fraction $f$ of the muons must decay in a 
channel that points in the desired 
direction. Muons live 100 times longer 
than charged pions. Since the decay fraction $f$ must be large we cannot use 
a linear muon decay channel unless we are prepared to build one that is 
tens of kilometers long. A more practical solution 
is to inject the muons into a storage ring with long straight sections. 
The useful decay fraction $f$ is just the length of the straight section 
divided by the circumference of the ring. 
It has been shown that $f \sim 0.3$ is achievable~\cite{design_study}. 
The resulting muon storage ring is sufficiently compact that it 
can be tilted downwards at a large angle so that the neutrino beam can 
pass through the Earth~\cite{nufact}, and very long baseline experiments 
($L \sim O(10^4)$~km) can be imagined.

Thus the ``neutrino factory'' concept~\cite{nufact,nufact_old} is 
to create a millimole/year muon source, rapidly accelerate the muons to the 
desired storage ring energy, and inject them into a storage ring 
with a long straight section that points in the desired direction. 
For discussion it is useful to define two types of neutrino factory: 
``entry--level'' and ``high--performance''. 
An entry--level neutrino factory~\cite{entry} 
can be thought of as a ``low'' intensity  
``low'' energy neutrino factory that we may (or may not) wish to build as a step 
towards the high--performance machine. We will take as typical parameters for an 
entry--level scenario a 20~GeV or 30~GeV storage ring delivering 
$O(10^{19})$ muon decays 
per year in the beam forming straight--section. With a 50~kt detector having 
a detection efficiency of 50\% an effective entry--level data sample would be 
$O(10^{21})$~kt--decays after a few years of running. 
Typical parameters for a 
high--performance scenario would be a 50~GeV ring delivering 
$O(10^{20})$ muon decays per year in the beam forming straight--section, 
yielding data samples $O(10^{22})$~kt--decays after a few years of running.

Neutrino factories would provide~\cite{nufact,physics_study}:
\begin{description}
\item{(i)} {\bf $\nu_e$ and $\overline{\nu}_e$ beams}, as well as 
$\nu_\mu$ and $\overline{\nu}_\mu$ beams !
\item{(ii)} {\bf High event rates.} With $2 \times 10^{20}$ muon 
decays per year in the beam--forming 
straight section of a 50~GeV neutrino factory the $\nu_\mu$ event rates in a distant 
detector would be about a factor of 60 higher than the corresponding rates for the 
next generation of conventional beams (NUMI at FNAL for example). 
These neutrino factory 
rates would yield tens of thousands of $\nu_\mu$ events per year within a reasonable 
sized detector on the other side of the Earth ($L \sim 10000$~km).
In addition, a near--detector a few hundred meters from the end of the beam--forming 
straight section of a 50~GeV neutrino factory would measure of the order of a million 
events per year per kg ! 
This fantastic rate would enable a revolution in non--oscillation 
neutrino experiments, which could be based on silicon pixel targets, 
polarized hydrogen targets, 
and detectors with fine segmentation and good particle identification. 
\item{(iii)} {\bf Narrow $\nu$ and $\overline{\nu}$ energy spectra.} 
Neutrinos from a neutrino factory have a much narrower 
energy spectrum than provided by a conventional ``wide--band'' beam. 
Hence, a neutrino factory beam 
can be thought of as being ``narrow band''.
\item{(iv)} {\bf Low systematic uncertainties.} Since the muon decay spectrum 
is very well known, the systematic uncertainties on the flux and spectrum of neutrinos 
at a distant experiment are expected to be significantly less than the corresponding 
uncertainties for a conventional beam. 
This would be expected to improve the ultimate precision of 
$\nu_\mu$ disappearance measurements.
\item{(v)} {\bf Polarization.} In the forward direction the $\nu_e$ flux at a neutrino 
factory is sensitive to the polarization of the muons in the storage ring. 
Hence, by controlling 
the polarization the $\nu_e$ component within the initial beam can be varied. 
In principle this
could be very useful, although a compelling case for muon polarization 
has yet to be demonstrated in a detailed analysis. 
\end{description}

Thus, compared with the next generation of conventional neutrino beams, 
neutrino factories offer the prospect of higher intensity neutrino and antineutrino 
beams containing $\nu_e$ as well as $\nu_\mu$, lower systematic uncertainties, a narrower beam 
energy distribution, and perhaps beam composition control via polarization. 
In addition the intensity increase would 
initiate a revolution in non--oscillation experiments. Its easy, therefore, to understand 
the current interest in neutrino factories.

\section{Neutrino oscillations}

Before we can 
discuss the physics potential of oscillation experiments at a neutrino factory 
we must first consider the theoretical framework used to describe neutrino oscillations. 
We know of three neutrino flavors: $\nu_e$, $\nu_\mu$, and $\nu_\tau$. 
Within the framework of three--neutrino oscillations, the flavor eigenstates are
related to the mass eigenstates by a $3\times3$ unitary matrix 
$U_{MNS}$~\cite{umns}:
\begin{equation}
\left(\begin{array}{c} \nu_e \\ \nu_\mu \\ \nu_\tau \end{array} \\ \right)=
\left(\begin{array}{ccc}
U_{e1} & U_{e2} & U_{e3} \\
U_{\mu1} & U_{\mu2} & U_{\mu3} \\
U_{\tau1} & U_{\tau2} & U_{\tau3} \\
\end{array}\right)
\left(\begin{array}{c} \nu_1 \\ \nu_2 \\ \nu_3 \end{array} \\ \right)
\; .
\end{equation}
In analogy with the CKM matrix, $U_{MNS}$ can be parameterized in terms of three mixing angles 
$\theta_{ij}$ and a complex phase $\delta$:
\begin{equation}
\left(\begin{array}{c} \nu_e \\ \nu_\mu \\ \nu_\tau \end{array} \\ \right)=
\left(\begin{array}{ccc}
c_{12}c_{13} & s_{12}c_{13} & s_{13}e^{-i\delta} \\
-s_{12}c_{23}-c_{12}s_{23}s_{13}e^{i\delta}
& c_{12}c_{23}-s_{12}s_{23}s_{13}e^{i\delta}
& s_{23}c_{13} \\
s_{12}s_{23}-c_{12}c_{23}s_{13}e^{i\delta}
& -c_{12}s_{23}-s_{12}c_{23}s_{13}e^{i\delta}
& c_{23}c_{13} \\
\end{array}\right)
\left(\begin{array}{c} \nu_1 \\ \nu_2 \\ \nu_3 \end{array} \\ \right)
\; ,
\end{equation}
where $c_{ij} = \cos\theta_{ij}$ and $s_{ij} = \sin\theta_{ij}$. If the
neutrinos are Majorana, there are two extra phases, but these do not
affect oscillations.
The evolution of the neutrino flavor states in vacuum is described by:
\begin{equation}
i \; \frac{d\nu_\alpha}{dt} \;=\; 
\sum_{\beta}{
\left(
\sum_{j}{
U_{\alpha j}U^\star_{\beta j} \frac{m^2_j}{2E_\nu}
}
\right)
\nu_\beta
} \; .
\end{equation}
Hence, the flavor oscillations are driven by the differences in the 
squares of the masses $m_j$. It is convenient to define: 
\begin{equation}
\Delta m^2_{ij} \equiv m^2_i - m^2_j \; .
\end{equation}
Oscillation probabilities depend upon the time--of--flight (and hence the 
baseline $L$), the $\Delta m^2_{ij}$, and $U_{MNS}$ (and hence 
$\theta_{12}, \theta_{23}, \theta_{13}$, and $\delta$).

\begin{figure}
\centering
\epsfxsize=6.in
\epsffile{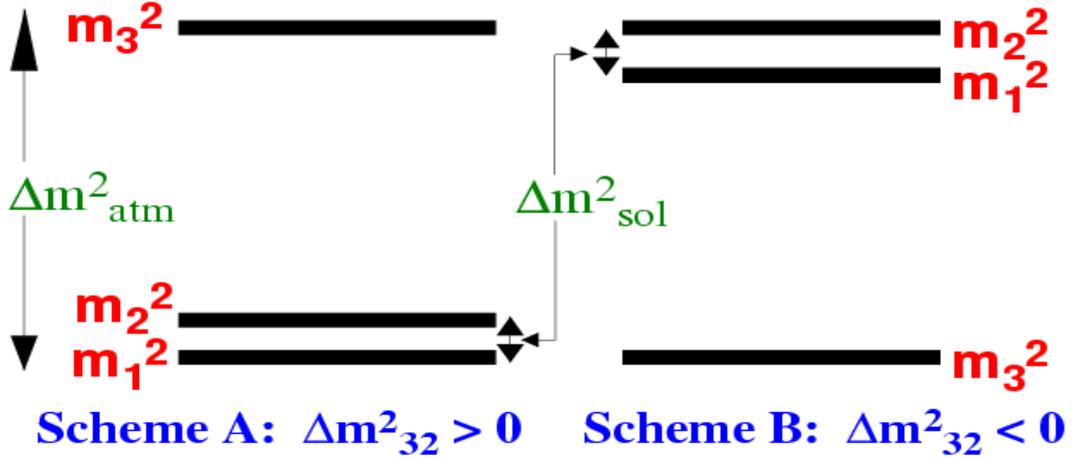}
\caption[]
{Alternative neutrino mass patterns that are consistent with neutrino 
oscillation explanations of the atmospheric and solar neutrino deficits.}
\label{fig1}
\end{figure}

The oscillation probabilities inferred from the 
atmospheric neutrino, solar neutrino, and LSND measurements 
can be used to constrain the oscillation parameters. 
For the time being we set aside the LSND oscillation results 
(which have not yet been confirmed by other experiments), and identify 
$\Delta m^2_{21}$ and 
$\Delta m^2_{32}$ as respectively the splittings that drive the solar and atmospheric 
neutrino oscillations. The atmospheric neutrino measurements imply that 
$|\Delta m^2_{32}| =$ (1.5 - 7)~$\times 10^{-3}$~eV$^2$ with an oscillation 
amplitude $\sin^2 2\theta_{atm} > 0.8$. 
There are four regions of parameter space consistent with the solar neutrino 
measurements: 
(a) MSW Small Mixing Angle (SMA): $|\Delta m^2_{21}| = (4 -10) \times 10^{-6}$~eV$^2$ 
with amplitude $\sin^2 2\theta_{sol} = 0.001 - 0.01$,
(b) MSW Large Mixing Angle (LMA): 
$|\Delta m^2_{21}| = (1.5 -10) \times 10^{-5}$~eV$^2$ 
with amplitude $\sin^2 2\theta_{sol} \sim 0.8$,
(c) MSW Long Wavelength (LOW): $|\Delta m^2_{21}| = (7 -20) \times 10^{-8}$~eV$^2$ 
with amplitude $\sin^2 2\theta_{sol} \sim 0.9$, and 
(d) Vacuum oscillations (VO): $|\Delta m^2_{21}| = (0.5 -8) \times 10^{-10}$~eV$^2$ 
with amplitude $\sin^2 2\theta_{sol} \sim 0.9$.
Recent preliminary solar neutrino results from SuperK seem to favor 
the LMA solution~\cite{ichep2000}, 
but it is perhaps too early to draw strong conclusions from this. 
In any event, it is evident that $|\Delta m^2_{21}| \ll |\Delta m^2_{32}|$.
However, we don't know whether $m_3$ is greater than or less than $m_2$, and 
hence there are two viable patterns for the neutrino mass spectrum (Fig.~\ref{fig1}).

How are neutrino oscillation measurements used 
to determine the oscillation parameters? 
To gain some insight it is useful to consider the oscillation probabilities 
$P(\nu_\alpha \rightarrow \nu_\beta)$  
using the approximation that oscillations driven by the 
small $\Delta m^2_{21}$ are neglected. This approximation is valid for 
long--baseline accelerator experiments. 
The resulting leading--oscillation probabilities for neutrinos 
of energy $E_\nu$~(GeV) 
propagating a distance $L$~(km) in vacuum  are~\cite{bpw80}
\begin{eqnarray}
P(\nu_e \rightarrow \nu_\mu) &=&
\sin^2\theta_{23} \sin^22\theta_{13} \sin^2(1.267 \Delta m^2_{32} L/E_\nu)
\; , \label{Pem}\\
P(\nu_e \rightarrow \nu_\tau) &=& 
\cos^2\theta_{23} \sin^22\theta_{13} \sin^2(1.267 \Delta m^2_{32} L/E_\nu)
\; , \label{Pet}\\
P(\nu_\mu \rightarrow \nu_\tau) &=&
\cos^4\theta_{13} \sin^22\theta_{23} \sin^2(1.267 \Delta m^2_{32} L/E_\nu)
\; . \label{Pmt}
\end{eqnarray}
The $L/E_\nu$ dependence of the oscillation probabilities can be used to 
determine $|\Delta m^2_{32}|$. However, 
the oscillating factors $\sin^2(1.267 \Delta m^2_{32} L/E_\nu)$ 
depend only on the magnitude of $\Delta m^2_{32}$ and not on its sign. Hence 
measurements of neutrino oscillations in vacuum cannot distinguish between 
the two viable mass eigenstate patterns shown in Fig.~\ref{fig1}. Fortunately 
the oscillation probabilities for transitions with a 
$\nu_e$ or $\overline{\nu}_e$ in the initial or final state  
do depend on the sign of $\Delta m^2_{32}$ if the neutrinos propagate through 
matter. We will return to this later. 
Note that the oscillation amplitudes in Eqs.~\ref{Pem}--\ref{Pmt} 
depend upon two mixing angles. It is clearly 
necessary to measure several oscillation modes to extract all of the 
mixing angles. 
Hence $\nu_e$ and $\overline{\nu}_e$ (as well as $\nu_\mu$ and 
$\overline{\nu}_\mu$ beams) are desirable. 

\section{What can we learn from $\nu$ oscillations ?}

The CHOOZ reactor ($\overline{\nu}_e$ disappearance) experiment places a limit 
on the $\overline{\nu}_e$ oscillation amplitude, yielding 
$\sin^22\theta_{13} < 0.1$. 
Interpreting the atmospheric neutrino results as evidence for 
$\nu_\mu \rightarrow \nu_\tau$ 
oscillations gives $\sin^2 2\theta_{23} \sim \sin^2 2\theta_{atm} > 0.8$. 
The solar neutrino measurements constrain 
$\sin^2 2\theta_{12} \sim  \sin^2 2\theta_{sol} \equiv 4|U_{e1}|^2|U_{e2}|^2$, 
to be $\sin^2 2\theta_{12} \sim$~0.8 -- 1 (LMA, LOW, VO) or 
$\sin^2 2\theta_{12} =$ 0.001 -- 0.01 (SMA). 
Hence we are on the 
threshold of measuring the three mixing angles, and learning 
something about the mixing matrix elements that govern neutrino oscillations.

This is exciting because 
there is a deep connection between the parameters that govern neutrino 
oscillations and physics at very high mass scales. The first clue to this 
connection comes from the smallness of the apparent neutrino masses. 
Direct limits on the electron neutrino mass from the tritium beta decay 
end point, 
together with cosmological constraints on the sum 
of the neutrino masses 
and the magnitude of the mass splittings obtained from the 
neutrino oscillation data,  
imply that all three neutrinos have masses $< 2$~eV, and are perhaps much 
smaller than this. If for example the masses are of 
the same order as the mass splittings, then the heaviest neutrino mass might 
be O(0.01--0.1)~eV. The well known seesaw mechanism~\cite{seesaw} 
provides a natural 
explanation for the smallness of these masses. If there exist right handed 
neutrinos $\nu_R$ (required by all GUT groups larger than SU(5)), and if 
lepton number is violated,  there will be both Dirac mass ($m_D$) terms 
and Majorana mass ($m_M$) terms in the Lagrangian. The seesaw mechanism then 
generates light neutrino masses of order $m^2_D / m_M$. With $m_D$ at the 
electroweak scale [$m_D \sim O(100$~GeV)] and $m_M$ at the 
Grand Unification scale 
($10^{15-16}$~GeV) neutrino masses in the desired range are natural.

Specific GUT models yield constraints on the neutrino mass eigenstates 
($m_1,m_2,m_3$), and predict the pattern of entries in the mass matrix $M$ 
(the so called ```texture'' of the mass matrix). 
The effective light neutrino mass matrix $M_\nu$ is related by the seesaw 
formula to the Dirac mass matrix $M_N$ (connecting $\nu_L$ and $\nu_R$) 
and the right--handed Majorana neutrino mass matrix 
$M_R$ (connecting $\nu_R$ and $\nu_R$):
\begin{equation}
M_\nu \; = \; -M^T_N M^{-1}_R M_N \;.
\end{equation}
The matrices $M_N$ and $M_R$ (and hence $M_\nu$) 
are predicted by GUT models in their corresponding flavor bases. 
The light neutrino masses are found by diagonalization of $M_\nu$, 
where the transformation matrix $U_\nu$ between the two bases is 
just $U_{MNS}$ :
\begin{equation}
U^\dag_{MNS} \; M_\nu \; U_{MNS} \; = \; {\rm diag}(m_1,m_2,m_3) \; ,
\end{equation}
with the charged lepton mass matrix diagonal in its flavor basis 
(more generally $U_{MNS} = U^\dag_LU_\nu$). 
Clearly in the lepton sector $U_{MNS}$ plays the role of the CKM matrix 
$V_{CKM}$ in the quark sector. 
Neutrino oscillation measurements constrain the pattern of the elements of $U_{MNS}$ 
and the pattern of the mass eigenstates $(m_1,m_2,m_3)$, and hence constrain the 
texture of the mass matrix $M_\nu$ which is predicted by GUT models.

We are familiar 
with the pattern of the CKM matrix elements, as parametrized 
by Wolfenstein~\cite{wolf}:
\begin{equation}
V_{CKM} \sim 
\left(\begin{array}{ccc}
1-\lambda^2/2 & \lambda & \lambda^3 \\
-\lambda & 1-\lambda^2/2 & \lambda^2 \\
\lambda^3 & \lambda^2 & 1 \\
\end{array}\right)
\; ,
\end{equation}
where $\lambda \simeq V_{us} \simeq 0.22$. 
The atmospheric neutrino measurements, 
which yield $\sin^2 2\theta_{23} > 0.8$, imply that $U_{\mu3}$ is large, 
and hence that $U_{MNS}$ has 
a different pattern to $V_{CKM}$.
Although this was not a priori expected, in recent months the large $\sin^2 2\theta_{23}$ 
has provoked a plethora of papers that demonstrate that large mixing in the (23) block 
of $U_{MNS}$ is not unnatural within the frameworks of a variety of specific 
GUT models. If $U_{MNS}$ can be completely determined by further oscillation 
measurements, the resulting constraints on the texture of $M_\nu$ 
will hopefully discriminate 
between GUT models (or maybe eliminate all of them). 
These same GUT models also predict proton decay rates 
and neutrinoless double beta decay rates. 
Hence, the presence or absence of proton decay and/or 
neutrinoless double beta decay can be used to further pin down the 
GUT alternatives.

Neutrino oscillation measurements offer a way of making a direct assault 
on our understanding of physics at high mass scales. With this in mind, it 
is difficult to think of neutrino experiments as merely a side--show to 
the high energy collider experimental program focussed on the origin of electroweak 
symmetry breaking. Rather, the neutrino oscillation program appears to be 
the corner stone of an attack on physics at the GUT scale. Over the 
next ten years the next 
generation of accelerator based neutrino oscillation experiments are 
expected to confirm the oscillation interpretation of the atmospheric 
neutrino deficit measurements, measure $\sin^2 2\theta_{23}$ and 
$|\Delta m^2_{32}|$ with precisions of about 10\%, and find evidence 
(via $\nu_\mu \rightarrow \nu_e$ oscillations) for a finite 
$\sin^2 2\theta_{13}$ if its value exceeds $\sim 0.01$. 
In addition, in the near future further solar 
neutrino measurements are expected to reduce the viable number of regions 
of parameter space to one (or none?). This progress will almost 
certainly keep neutrino oscillations in the limelight for the coming 
decade. However, to discriminate between GUT models (or point the way 
to an alternative theoretical framework) we will want to know more. 
In particular we will want to be sure we have the right oscillation 
framework (three--flavor?), precisely measure (or put stringent limits on) all 
of the $U_{MNS}$ elements, 
and determine whether there is 
significant CP violation in the lepton sector. 

\section{Which measurements are important ?}

Once $U_{MNS}$ has been measured 
we may find that it conforms to some recognizable pattern. 
With what precision will we want to measure $U_{MNS}$? 
We will clearly want to know which elements are approximately 
0 or 1. Since GUT predictions have uncertainties associated 
with the evolution from high mass scales to low mass scales, 
the difference between 0 and some sufficiently small number $\epsilon$, or 
between 1 and $(1-\epsilon)$, is unlikely to discriminate between 
GUTs. Lacking any guidance for the size of the GUT uncertainties $\epsilon$ 
we will recklessly seek guidance from $V_{CKM}$ on the required precision 
with which we want to know $U_{MNS}$. Noting that some elements of 
$V_{CKM}$ differ from unity by as little as O(0.01) and some elements 
differ from 0 by as little as 
O(0.01), we are motivated to measure all elements of $U_{MNS}$ 
with a precision O(0.01). 
With this goal in mind, in ten years time the big neutrino oscillation GUT 
questions 
that will need to be answered to pin down $U_{MNS}$ and the 
pattern of neutrino masses, and hence discriminate between GUT models, 
are likely to be:
\begin{description}
\item{{\bf (Q1)}} If $\nu_\mu \rightarrow \nu_e$ has not been observed, then 
how small is $\sin^2 2\theta_{13}$ ? Is it $O(10^{-2})$~? Is it smaller than $10^{-3}$~? 
If  $\nu_\mu \rightarrow \nu_e$ has been observed, then 
precisely how big ($\pm 10$\%) is $\sin^2 2\theta_{13}$ ? 
\item{{\bf (Q2)}} What is the pattern of neutrino masses (Fig.~\ref{fig1} 
scheme A or scheme B)~?
\item{{\bf (Q3)}} Is there CP violation in the lepton sector, 
and how big is the phase $\delta$~?
\item{{\bf (Q4)}} How close ($\pm$~few~\%) is $\sin^2 2\theta_{23}$ to 1~?
\item{{\bf (Q5)}} If we are left with the SMA solar solution, then 
precisely how big ($\pm 10$\%) is $\sin^2 2\theta_{12}$~? 
If we are left with the LMA, LOW, or VO solar solutions, then 
how close ($\pm$~few~\%) is $\sin^2 2\theta_{12}$ to 1~?
\item{{\bf (Q6)}} Do neutrino oscillations involve only 3 flavors, 
or are there light sterile neutrinos~? If in a few years 
time the totality 
of the solar, atmospheric, and accelerator data suggests the participation 
of sterile neutrinos, this question goes to the top of the list.
\end{description}
The following describes how experiments at a neutrino 
factory can answer these questions.

\begin{figure}
\centering
\leavevmode
\epsfxsize=6.in
\hspace{-1.0cm} \epsffile{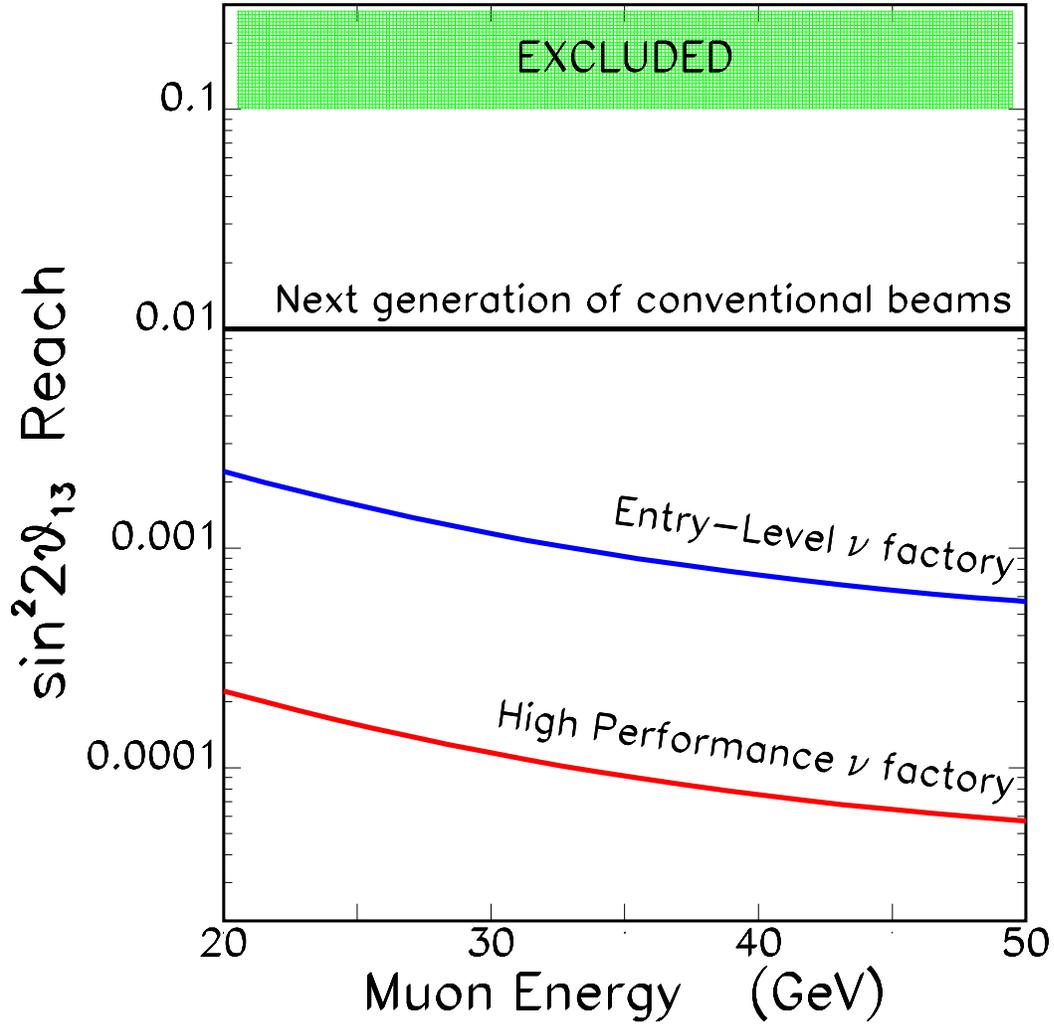}
\vspace{1.cm}
\caption[]
{Limiting $\sin^2 2\theta_{13}$ sensitivity for the observation 
of $\nu_\mu \rightarrow \nu_e$ oscillations expected at the next 
generation of conventional neutrino beams compared, as a function 
of muon energy, with the corresponding sensitivities for the 
observation of $\nu_e \rightarrow \nu_\mu$ oscillations at 
entry-level ($10^{21}$~kt--decays) and high--performance 
($10^{22}$~kt-decays) neutrino factories. The neutrino factory calculations 
are for $L = 2800$~km, $\Delta m^2_{32} = 0.0035$~eV$^2$, and 
$\sin^2 2\theta_{23} = 1$. Figure based on calculations presented 
in Ref.~\cite{entry}.}
\label{fig3}
\end{figure}

\section{Determining $\sin^2 2\theta_{13}$}

The next generation of long baseline accelerator 
experiments~\cite{k2k,minos,icanoe,opera} are 
expected to observe $\nu_\mu \rightarrow \nu_e$ if 
$\sin^2 2\theta_{13} > 0.01$, about an order of magnitude below the
presently excluded region. If $\sin^2 2\theta_{13}$ is smaller than 
this, then $|U_{e3}| < 0.05$. The question will then be, is $U_{e3}$ 
just small, or is it very small [$|U_{e3}| < O(0.01)$ in which case 
$\sin^22\theta_{13} \sim 4U^2_{e3} \le O(10^{-4})$]~? 
To address this question
we would need to improve the $\sin^2 2\theta_{13}$ sensitivity 
by about two orders of magnitude. Hence we would like to be able 
to observe $\nu_\mu \rightarrow \nu_e$ or $\nu_e \rightarrow \nu_\mu$
oscillations if $\sin^2 2\theta_{13} > 0.0001$

At a neutrino factory $\nu_e \rightarrow \nu_\mu$ oscillations are 
the preferred mode for probing small $\sin^2 2\theta_{13}$.
Consider a neutrino 
factory in which positive muons are stored. The initial neutrino beam 
contains $\overline{\nu}_\mu$ and $\nu_e$. In the absence of oscillations 
charged current (CC) interactions of the $\overline{\nu}_\mu$ in 
a far detector will produce positive muons, i.e. muons of the same 
sign as those stored in the ring. In the presence of 
$\nu_e \rightarrow \nu_\mu$ oscillations there will also be 
$\nu_\mu$ CC interactions in the detector, producing negative 
muons, i.e. muons of opposite charge to those stored in the ring. 
Hence, the experimental signature for $\nu_e \rightarrow \nu_\mu$ 
oscillations is the appearance of ``wrong--sign'' muons. 

In a long baseline neutrino factory experiment the expected number of 
wrong--sign muon events will depend on the oscillation amplitude which, 
to a good approximation, is proportional to $\sin^2 2\theta_{13}$ 
(Eq.~\ref{Pem}). The other relevant oscillation parameters 
($\sin^2 2\theta_{23}$ and $|\Delta m^2_{32}|$) will be known ($\pm 10$\%)  
after the next generation of accelerator based experiments. 
The  $\sin^2 2\theta_{13}$ sensitivity 
depends upon the number of muons that have decayed in the beam--forming
straight section $N_{dec}$, the muon energy $E_\mu$, the baseline $L$, 
the detector mass $M_{det}$, and the detector efficiency, resolutions, 
and backgrounds. 
Detailed simulations that include detector efficiencies, 
resolutions, and backgrounds, have explored the sensitivity 
as a function of $N_{dec} \times M_{det}$, $E_\mu$, and $L$. 

To illustrate the anticipated limiting $\sin^2 2\theta_{13}$ sensitivity 
at a neutrino factory 
consider a 30~GeV muon storage ring pointing at a detector at $L = 7400$~km, 
and let  $N_{dec} \times M_{det} = 2 \times 10^{21}$~kt--decays 
(corresponding to an entry--level scenario).  
It has been shown~\cite{camp} that, for values of 
$|\Delta m^2_{32}|$ in the center of the preferred SuperK range, 
the absence of a wrong--sign muon signal in 
this entry--level scenario would result in an upper limit on 
$\sin^2 2\theta_{13}$ of a few~$\times 10^{-3}$. 
Similar results have been obtained for $L = 2800$~km~\cite{entry}.  
The limiting sensitivity is shown for $L = 2800$~km 
as a function of neutrino 
factory energy in Fig.~\ref{fig3}. 
The limiting $\sin^22\theta_{13}$ 
sensitivity at a 30~GeV 
neutrino factory delivering a factor of 10 more muon decays/year 
improves to better than $2 \times 10^{-4}$. 
At a high performance 50~GeV neutrino factory 
the limiting $\sin^2 2\theta_{13}$ 
sensitivity would be better than $10^{-4}$~\cite{cervera}.

We conclude that, if no $\nu_\mu \rightarrow \nu_e$ signal is 
observed by the next generation of long baseline experiments, and 
therefore $|U_{e3}| < O(0.05)$, 
a search for $\nu_e \rightarrow \nu_\mu$ oscillations at an entry--level 
neutrino factory would facilitate an order of 
magnitude improvement in the sensitivity to a finite $\sin^22\theta_{13}$, 
probing $\sin^2 2\theta_{13}$ at the $10^{-3}$ level. 
Hence, the entry--level experiment would either 
make a first observation of $\nu_e \rightarrow \nu_\mu$ oscillations 
or significantly improve 
the limits on $|U_{e3}|$. In either case we would want to upgrade the 
performance of 
the neutrino factory to precisely measure, or probe smaller values of, 
$\sin^2 2\theta_{13}$. 
A 50~GeV high performance neutrino factory would enable 
$\nu_e \rightarrow \nu_\mu$ oscillations to be observed 
for values of $\sin^22\theta_{13}$ as small as $10^{-4}$.

What if the next generation of long baseline accelerator experiments observes 
a $\nu_\mu \rightarrow \nu_e$ signal ?  In this case, $\sin^2 2\theta_{13} > 0.01$, 
and depending on its exact value the experiments would be expected to have observed 
from a few to a few tens of signal events. The question then becomes, what is the 
precise value of $\sin^2 2\theta_{13}$ ($\pm 10$\%)?  To address this question 
will require O(100) signal events (or more if there is significant background). 
An entry--level neutrino factory, providing an order of magnitude improvement 
in sensitivity with negligible backgrounds, 
would be expected to determine $\sin^2 2\theta_{13}$ with the 
desired precision, and would be able to exploit the substantial signal 
to determine the pattern of neutrino masses~!

\begin{figure}
\centering
\leavevmode
\epsfxsize=6.in
\hspace{-1.0cm} \epsffile{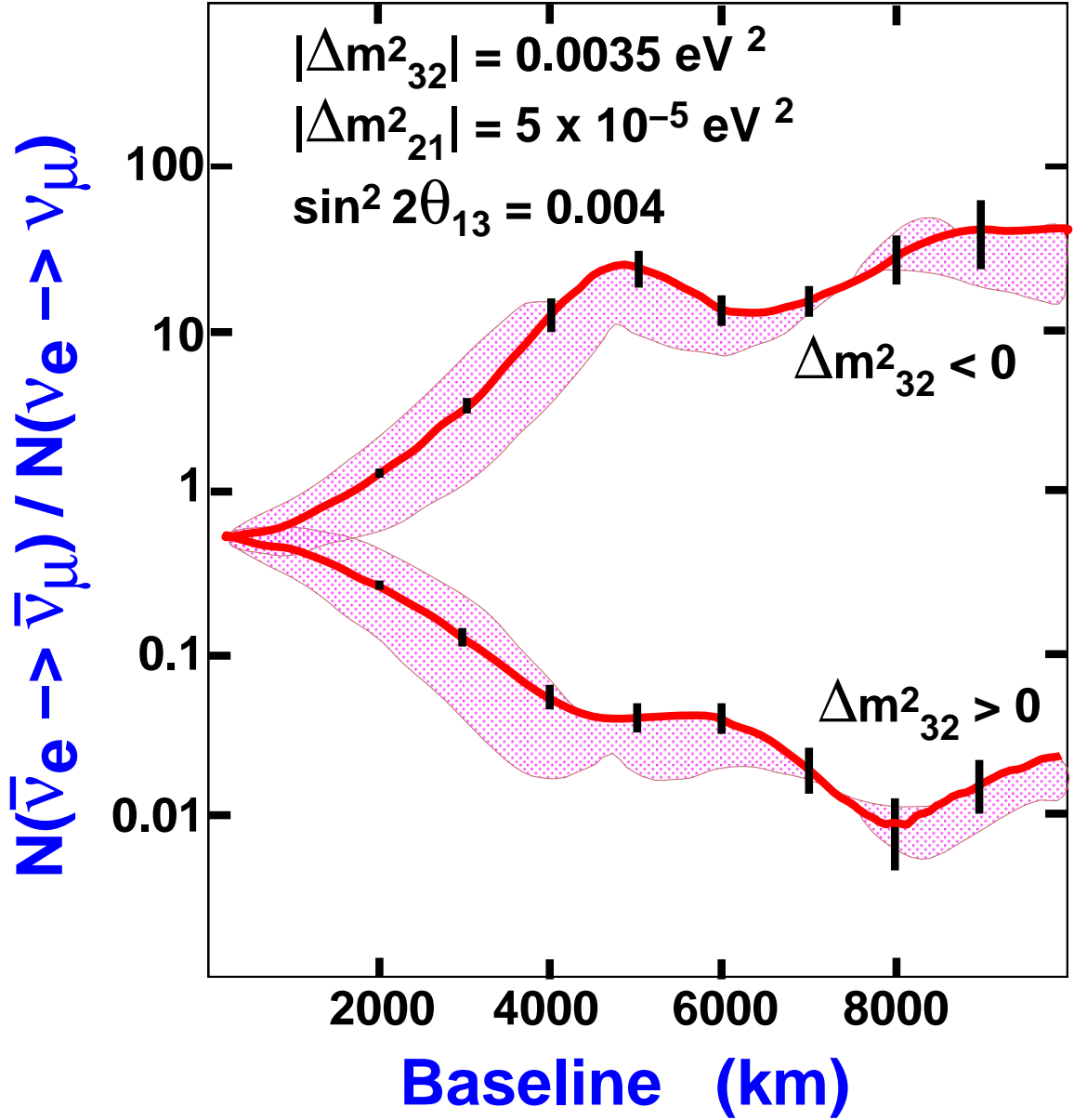}
\vspace{1.cm}
\caption[]
{Predicted ratios of wrong--sign muon event rates when positive and negative 
muons are stored in a 20~GeV neutrino factory, shown as a function of 
baseline. A muon measurement threshold of 4~GeV is assumed. 
The lower and upper bands correspond respectively 
to schemes A and B in Fig.~1. The 
widths of the bands show how the predictions vary as the CP violating 
phase $\delta$ is varied from $-\pi/2$ to $+\pi/2$, with the thick lines 
showing the predictions for $\delta = 0$. The statistical error bars 
correspond to a high--performance neutrino factory yielding a 
data sample of $10^{21}$ decays with a 50~kt detector. 
Figure based on calculations presented 
in Ref.~\cite{entry}.}
\label{fig2}
\end{figure}

\section{The pattern of neutrino masses}

How can we distinguish between the two mass splitting patterns 
in Fig.~1? Fortunately the oscillation probabilities for transitions 
involving a $\nu_e$ or $\overline{\nu}_e$ are modified if the neutrinos 
propagate through matter, and the modification depends upon the 
sign of $\Delta m^2_{32}$~\cite{parke,bgrw2}. 

In the leading oscillation
approximation the probability for $\nu_e\to\nu_\mu$ 
oscillations in matter of constant density $\rho(x)$ and electron fraction 
$Y_e(x)$, is given by:
\begin{equation}
P(\nu_e\to\nu_\mu) = s_{23}^2 \sin^22\theta_{13}^m \sin^2\Delta_{32}^m\,,
\label{eq:Pemu}
\end{equation}
where
\begin{equation}
\sin^22\theta_{13}^m =
{ \sin^22\theta_{13}\over \left( {A\over\Delta m^2_{32}} -
\cos 2\theta_{13} \right)^2 + \sin^22\theta_{13} }
\label{eq:amp13m}
\end{equation}
and
\begin{equation}
\Delta_{32}^m = { 1.27 \Delta m_{32}^2 \,({\rm eV^2})\,L\,({\rm km)} \over
E_\nu\,\rm(GeV) }\; \sqrt{ \left( {A\over\Delta m^2_{32}} - \cos2\theta_{13}
\right)^2 + \sin^22\theta_{13}} \,,
\label{eq:delta32m}
\end{equation}
and $A$ is the matter amplitude:
\begin{equation}
A = 2\sqrt2\, G_F\, Y_e\rho E_\nu = 1.52 \times 10^{-4}{\rm\,eV^2}\, Y_e\,
\rho\,({\rm g/cm^3}) E_\nu\,({\rm GeV}) \,.
\label{eq:A}
\end{equation}
For $\bar\nu_e\to\bar\nu_\mu$ oscillations, the sign of $A$ is reversed
in Eqs.~(\ref{eq:amp13m}) and (\ref{eq:delta32m}). For
$\sin^22\theta_{13} \ll 1$ and $A \sim \Delta m^2_{32} > 0$ ($-A \sim
\Delta m^2_{32} < 0$), $P(\nu_e\to\nu_\mu)$ is enhanced (suppressed) and
$P(\bar\nu_e\to\bar\nu_\mu)$ is suppressed (enhanced) by matter
effects. Thus a comparison of the $\nu_e\to\nu_\mu$ CC rate with the 
$\bar\nu_e\to\bar\nu_\mu$ CC rate discriminates between the two signs of
$\Delta m^2_{32}$.

To illustrate how well the sign of $\Delta m^2_{32}$ can be determined 
at a neutrino factory, consider an experiment downstream of 
a 20~GeV neutrino factory. 
Let half of the 
data taking be with $\mu^+$ stored, and the other half with $\mu^-$ 
stored. In Fig.~\ref{fig2} the predicted ratio of wrong sign muon events 
$R \equiv N(\overline{\nu}_e \rightarrow \overline{\nu}_\mu) / 
N(\nu_e \rightarrow \nu_\mu)$ is shown as a function of baseline for
$\Delta m^2_{32} = +0.0035$~eV$^2$ and $- 0.0035$~eV$^2$, with 
$\sin^2 2\theta_{13}$ set to the small value 0.004. 
The figure shows two bands. The 
upper (lower) band corresponds to $\Delta m^2_{32} < 0$ ($> 0$). 
Within the bands the CP phase $\delta$ is varying (more on this later). 
At short baselines the bands converge, and the ratio $R = 0.5$ since 
the antineutrino CC cross-section is half of the neutrino CC cross-section. 
At large distances matter effects enhance $R$ if $\Delta m^2 > 0$ and 
reduce $R$ if $\Delta m^2 < 0$, and the bands diverge. Matter effects 
become significant for $L$ exceeding about 2000~km. The error bars indicate 
the expected statistical uncertainty on the measured $R$ 
with a data sample of $5 \times 10^{22}$~kt-decays. 
With these statistics, 
the sign of $\Delta m^2_{32}$ is determined with very high statistical significance.
With an order of magnitude smaller data sample (entry level scenario) or with 
an order of magnitude smaller $\sin^2 2\theta_{13}$ the statistical uncertainties 
would be $\sqrt{10}$ larger, but the 
sign of $\Delta m^2_{32}$ could still be determined with convincing precision 
in a long baseline experiment. 

A more detailed analysis~\cite{bgrw3} has shown that the 
pattern of neutrino masses could be determined at 
a 20~GeV neutrino factory delivering a few times $10^{19}$ ($10^{20}$) decays 
per year provided $\sin^2 2\theta_{13} > 0.01$ ($0.001$). This `$\sin^2 2\theta_{13}$ 
``reach'' improves with neutrino factory energy ($\sim E^{3/2}_\mu$), and a higher 
energy neutrino factory could therefore probe the mass pattern for $\sin^2 2\theta_{13}$ 
smaller than 0.001.

\section{CP violation in the lepton sector}

The oscillation probabilities $P(\nu_\alpha \rightarrow \nu_\beta)$ can 
be written in terms of CP--even and CP--odd contributions:
\begin{equation}
P(\nu_\alpha \rightarrow \nu_\beta) \; = \; 
P_{\rm CP-even}(\nu_\alpha \rightarrow \nu_\beta) 
		+ P_{\rm CP-odd}(\nu_\alpha \rightarrow \nu_\beta) \; ,
\end{equation}
where
\begin{equation}
\begin{array}{rl}
	P_{\rm CP-even}(\nu_\alpha \rightarrow \nu_\beta) =&P_{\rm CP-even}(
		\bar{\nu}_\alpha \rightarrow \bar{\nu}_\beta)\\[0.1in]
  	=&\delta_{\alpha\beta} -4\sum_{i>j}\ Re\ (U_{\alpha i}
		U^*_{\beta i}U^*_{\alpha j}U_{\beta j})\sin^2 
		({{\Delta m^2_{ij}L}\over{4E_\nu}})\\[0.1in]
	P_{\rm CP-odd}(\nu_\alpha \rightarrow \nu_\beta) =&-P_{\rm CP-odd}(
		\bar{\nu}_\alpha \rightarrow \bar{\nu}_\beta)\\[0.1in]
        =&2\sum_{i>j}\ Im\ (U_{\alpha i}U^*_{\beta i}U^*_{\alpha j}
          U_{\beta j})\sin ({{\Delta m^2_{ij}L}\over{2E_\nu}})\\[0.1in]
\end{array}
\label{cprels}
\end{equation}

Hence, if there is CP violation in the lepton sector 
it might be observable at a neutrino factory~\cite{derujula} 
by comparing 
$\nu_e \rightarrow \nu_\mu$ with $\overline{\nu}_e \rightarrow \overline{\nu}_\mu$ 
probabilities, which we have seen can be done by measuring 
wrong--sign muon production when respectively $\mu^+$ and $\mu^-$ are 
stored (Fig.~\ref{fig2}). However, CP violation requires that (at least) two 
mass splittings contribute to the oscillations. This is why the sensitivity 
to CP violation (shown by the bands in Fig.~\ref{fig2}) vanishes at $L \sim 7000$~km, 
the distance at which $\sin^2(1.267\Delta m^2_{32} L/E_\nu) \rightarrow 0$ for 
neutrinos from a 20~GeV storage ring.
The baseline must be chosen carefully!  
The modification to $R$ also becomes harder to measure in a long baseline experiment 
as the contribution from the sub--leading scale decreases 
(i.e. for small $\Delta m^2_{21}$ or small oscillation amplitude). 
Within the 
framework of three--flavor oscillations with the two $\Delta m^2$ 
scales defined by the atmospheric and solar neutrino deficits, 
CP violation is only likely to be observable at a neutrino factory if 
the LMA solar solution defines the correct region of parameter space 
and $|\Delta m^2_{21}|$ is not too small. 
Interestingly, the LMA solution seems to be favored by the most recent 
SuperK data, but we must wait a little longer to see whether this is
confirmed. Finally, to have an observable CP violating rate 
$P(\nu_e \rightarrow \nu_\mu)$ must not be too small, which means 
that $\sin^2 2\theta_{13}$ must not be too small. 

In the example 
shown in Fig.~\ref{fig2}, with $\sin^2 2\theta_{13} = 0.004$, it is 
apparent that if $L$ is chosen to be 3000--4000~km, 
the predicted ratio $R$ varies significantly  
as the value of $\delta$ varies from 0 to $\pm \pi/2$. 
We might therefore suspect that with this value of $\sin^2 2\theta_{13}$ a 
high--performance neutrino factory would enable us to observe 
CP violation and determine $\delta$. However, before we can conclude this 
we must consider backgrounds and systematics, including the correlations 
between the fitted oscillation parameters that arise when all  
parameters are allowed to vary. Fortunately detailed studies have 
been made~\cite{cervera}, 
including backgrounds and global fits to all of the 
observed neutrino and antineutrino distributions. For $\sin^2 2\theta_{13}$ 
as small as 0.005, a 50~GeV high--performance neutrino factory could 
distinguish $\delta = 0$ from $\pi/2$ provided 
$|\Delta m^2_{21}| > 2 \times 10^{-5}$~eV$^2$. 
With larger $|\Delta m^2_{21}|$ a reasonable measurement of $\delta$ can 
be made ($\sigma_\delta \sim \pm 15^\circ$ if $\Delta m^2_{21} = 1 \times 10^{-4}$~eV$^2$ 
and $\sin^2 2\theta_{13} = 0.005$, for example).

We conclude that if the LMA solution turns out to be the correct solution 
to the solar neutrino deficit problem and 
$|\Delta m^2_{21}| > 2 \times 10^{-5}$ 
then CP violation would be observable at a high performance neutrino factory 
provided $\sin^2 2\theta_{13}$ is larger than $\sim 0.005$.

\section{Precise measurement of $\sin^22\theta_{23}$ and $|\Delta m^2_{32}|$}

How close is $\sin^2 2\theta_{23}$ to 1 ? 
In a long baseline experiment using a $\nu_\mu$ beam, if the baseline 
$L$ is close to the first oscillation maximum, 
the oscillations $\nu_\mu \rightarrow \nu_x$ will produce 
a dip in the observed 
$\nu_\mu$ spectrum. The position of the dip  
is determined by $|\Delta m^2_{32}|$, and its depth is 
determined by $\sin^2 2\theta_{23}$. A fit to the spectra measured by 
the next generation of long baseline accelerator experiments, 
with $L = 730$~km, 
is expected to yield $\Delta m^2_{32}$ and 
$\sin^2 2\theta_{23}$ with precisions of about 10\%~\cite{minos,icanoe}. 

It has been shown~\cite{camp} that a few years of running at an 
entry--level 30~GeV neutrino factory with $L = 7400$~km would 
(i) yield a comparable statistical precision on the 
determination of $\sin^2 2\theta_{23}$, with a smaller systematic uncertainty 
arising from the uncertainty on the neutrino flux, and 
(ii) improve the precision 
on $|\Delta m^2_{32}|$ to about 1\%. A high-performance 30~GeV 
neutrino factory would enable $\sin^2 2\theta_{23}$ to be measured with a 
precision of about 5\%. 
A systematic study to optimize $L$ and $E_\mu$ for 
these measurements has not been performed, and hence it may be possible to 
improve on these precisions with optimal choices.

\section{Determining $\sin^2 2\theta_{12}$ and $|\Delta m^2_{21}|$}

It will be a challenge to directly measure 
the sub--leading oscillation parameters 
$\sin^2 2\theta_{12}$ and $|\Delta m^2_{21}|$ 
in long--baseline accelerator experiments since 
the associated oscillation probabilities tend to be very small. 
For example, with $\Delta m^2_{21} = 10^{-5}$~eV$^2$, 
$L = 10^4$~km, and $E_\nu = 1$~GeV, the oscillating factor 
in the transition probabilities is given by  
$\sin^2(1.267 \Delta m^2 L / E_\nu) = 0.016$. 
Hence oscillations driven by the leading  
scale ($\Delta m^2_{32}$) tend to dominate unless the 
associated amplitude is very small. This could be the 
case for $\nu_e \rightarrow \nu_\mu$ oscillations if 
$\sin^2 2\theta_{13}$ is very small or zero. 
As an example, with $\sin^2 2\theta_{13} = 0$, 
$\sin^2 2\theta_{12} = 0.8$, and 
$\Delta m^2_{21} = 5\times 10^{-5}$~eV$^2$, 
it has been shown~\cite{entry} that $\nu_e \rightarrow \nu_\mu$ 
oscillations might be observed at a high performance 
neutrino factory with $L \sim 3000$~km, but would 
require background levels to be no larger than 
$O(10^{-5})$ of the total CC rate. If 
$\sin^2 2\theta_{12} << 1$ or $\Delta m^2_{21} < 10^{-5}$~eV$^2$ 
the oscillation rate would appear to be too low to observe 
even at a high--performance neutrino factory.

We conclude that, within the framework of three flavor oscillations 
that give rise to the atmospheric and solar neutrino deficits, 
direct observation of oscillations driven by the 
sub--leading scale, and hence direct measurement of 
$\sin^2 2\theta_{12}$ and $|\Delta m^2_{21}|$, 
might be feasible at a high--performance neutrino factory, 
but only if the LMA solution correctly describes the 
solar neutrino deficit and $\sin^2 2\theta_{13}$ is zero 
(or very small).

\section{The potential for surprises}

So far we have considered only three--neutrino oscillations 
with the $\Delta m^2_{ij}$ chosen to 
account for the solar and atmospheric neutrino deficits. 
What if:
\begin{description}
\item{(i)} The LSND oscillation results are confirmed?
\item{(ii)} The solar neutrino deficit has nothing to do with oscillations?
\item{(iii)} There are more than three flavors participating in the 
oscillations (light sterile neutrinos)?
\item{(iv)} Neutrino oscillation dominates the solar and atmospheric 
deficit results, but is not the whole story (e.g. neutrino decay, ... )?
\end{description}
Although it is tempting to apply Occam's razor, and neglect  
these exciting possibilities, we must remember that neutrino 
oscillations require physics beyond the Standard Model, and  
we might be in for some surprises.

The best way of ensuring that we have the right oscillation framework, 
and are not missing any additional new physics, is to measure, 
as a function of $L/E_\nu$, 
all of the oscillation modes (appearance and disappearance, neutrinos and 
antineutrinos) that we can, and then check for overall consistency of 
the oscillation parameters. With a conventional neutrino beam the 
modes that can in principle be measured are 
(a) $\nu_\mu$ disappearance,
(b) $\overline{\nu}_\mu$ disappearance, 
(c) $\nu_\mu \rightarrow \nu_\tau$,
(d) $\overline{\nu}_\mu \rightarrow \overline{\nu}_\tau$
(e) $\nu_\mu \rightarrow \nu_e$, and 
(f) $\overline{\nu}_\mu \rightarrow \overline{\nu}_e$. 
At a neutrino factory all of these can be measured, plus the 
additional modes: 
(g) $\nu_e$ disappearance, 
(h) $\overline{\nu}_e$ disappearance, 
(i) $\nu_e \rightarrow \nu_\tau$,
(j) $\overline{\nu}_e \rightarrow \overline{\nu}_\tau$
(k) $\nu_e \rightarrow \nu_\mu$, and 
(l) $\overline{\nu}_e \rightarrow \overline{\nu}_\mu$.

To illustrate the power of the additional measurements at a 
neutrino factory we consider one example: 
suppose the LSND oscillation results have been confirmed 
and we wish to discriminate between three--neutrino oscillations 
(describing the LSND and atmospheric results, with the solar 
neutrino deficit due to something else) or 
four--neutrino 
oscillations with three active flavors and one 
sterile neutrino (describing LSND, atmospheric, and solar neutrino 
results). It has been shown that~\cite{bgrw4} in the four--neutrino case, 
if three-neutrino oscillations 
are incorrectly assumed the parameters $\sin^2 2\theta_{12}$, 
$\sin^2 2\theta_{13}$ and $\delta$ determined by short baseline 
$\nu_e \rightarrow \nu_\tau$ and $\nu_e \rightarrow \nu_\mu$ 
measurements at a neutrino factory would be inconsistent 
with the same parameters determined from $\nu_\mu \rightarrow \nu_\tau$ 
measurements. Hence, the additional oscillation modes that can be 
probed at a neutrino factory offer discrimination between different 
hypotheses. Note that in both these three--flavor and four--flavor 
cases CP violation might be observable in both $\nu_e \rightarrow \nu_\mu$ 
and $\nu_\mu \rightarrow \nu_\tau$ oscillations 
at a short baseline neutrino factory experiment,  
even at an entry level neutrino facility~!

In general any surprise that forces us to depart from the 
minimal three--neutrino 
(solar plus atmospheric neutrino deficit) oscillation framework 
will increase the need to explicitly measure (or place 
stringent limits on) all of the 
appearance and disappearance channels. Hence surprises are  
likely to strengthen, rather than weaken, the already strong case for a 
neutrino factory!

\section{Questions about staging}

We hope that in a few years time the R\&D needed for a  
neutrino factory will be complete. What neutrino--factory should 
we then propose? The physics potentials for entry--level and high--performance 
neutrino factories are compared in Table~1 
with the corresponding potential for the
next generation of long--baseline oscillation experiments.
There appears 
to be a strong physics case for a high--performance facility that can 
deliver a few $\times 10^{20}$ useful muon decays per year. 
If we believe we can obtain the required resources for this, and can 
build a high performance factory without first building a more modest 
facility to climb the technical learning curve, then that is what 
we should propose to do. 
However, cost and/or technical considerations 
may make staging necessary.

\begin{table}
\caption{\label{compare_tab}
Comparison of oscillation physics measurements at the next 
generation of conventional accelerator based long-baseline 
experiments with the corresponding programs at entry-level 
and high--performance neutrino factories.
}
\begin{center}
\begin{tabular}{l|ccc}
\hline \hline
 &Next Generation&Entry&High\\
 &Conventional&Level&Performance\\
\hline
$\sin^22\theta_{13}$ reach   &0.01&$10^{-3}$&$10^{-4}$\\
\hline
$\Delta m^2_{32}$ sign       & NO &if $\sin^22\theta_{13}>0.01$&if $\sin^22\theta_{13}>0.001$\\
\hline
CP Violation                 & NO & NO &if $\sin^22\theta_{13}>0.005$ and\\
                             &    &    &$|\Delta m^2_{21}| > 2\times10^{-5}$ eV$^2$\\
\hline
$\sin^22\theta_{23}$ precision&10\%&10\%&$<$5\%\\
\hline
$|\Delta m^2_{32}|$ precision &10\%&10\%&$<$1\%\\
\hline
sub-leading                  & NO & NO & if LMA and \\
oscillations ?               &    &    &$\sin^22\theta_{13}\le$ few $\times 10^{-5}$\\
\hline\hline
\end{tabular}
\end{center}
\end{table}

\begin{figure}
\centering
\vspace{-2.0cm}
\epsfxsize=6.5in
\epsffile{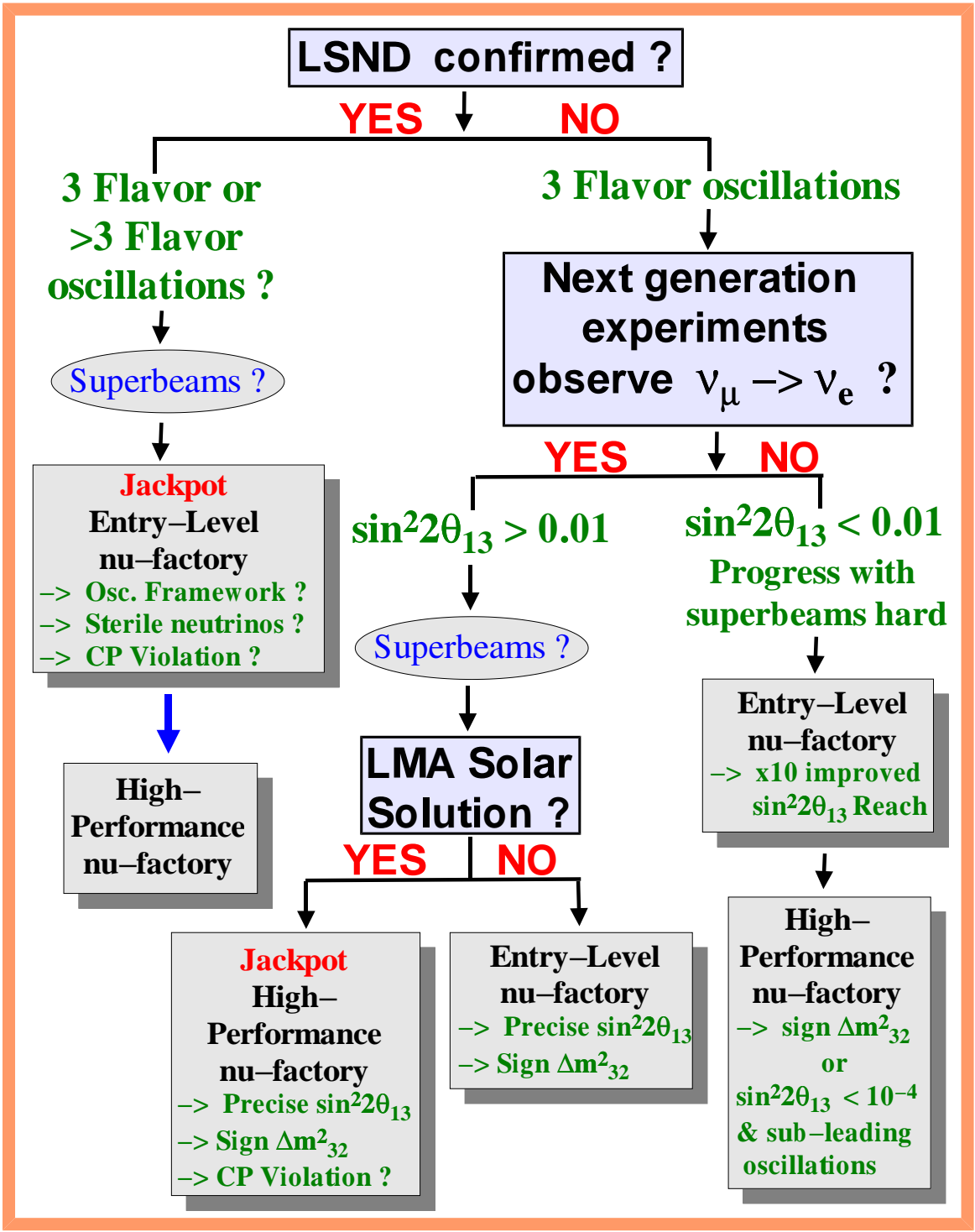}
\vspace{-2.cm}
\caption[]
{Staging scenario decision tree; a first attempt.
}
\label{fig4}
\end{figure}

Fortunately there are a variety of possible staging options. As an 
intermediate step towards a high--performance neutrino factory we 
can consider (i) a proton accelerator system of the type needed for 
a neutrino factory, but used to drive a conventional 
neutrino ``superbeam'', or 
(ii) an entry--level neutrino factory. The superbeam 
facility might also include the neutrino factory target station and 
pion decay channel, providing an intense stopped muon source and, 
downstream of the decay channel, an intense low energy neutrino beam.

The pros and cons for any given staging strategy will depend upon 
the results from the next generation of neutrino oscillation 
experiments. This dependence is illustrated in Fig.~\ref{fig4} which 
shows a first ``strawman'' attempt at constructing a physics scenario 
dependent decision tree. If the LSND oscillation results are confirmed 
by the MiniBooNE experiment~\cite{miniboone} the immediate big questions are 
likely to be: What is the 
oscillation framework? Do light sterile neutrinos participate in the 
oscillations? Is there significant CP violation in the lepton sector?
These questions can be addressed by an entry--level neutrino factory, 
and hence if LSND results are confirmed then neutrino factories will 
hit the physics jackpot! Beyond an entry--level facility there will 
probably be so much to measure and sort out that a high--performance 
factory would be desired. 
Superbeams might also be proposed to try to make some 
progress even before a neutrino factory could be built, although 
whether these high intensity conventional beams can address any of the 
central questions requires further study. 

If the LSND oscillation results are not confirmed it seems likely  
that three--flavor oscillations will be accepted as the right phenomenological 
framework. 
In this case, the preferred staging strategy will probably depend upon 
whether the next generation of long baseline accelerator experiments  
observe, or do not observe, $\nu_\mu \rightarrow \nu_e$ oscillations, 
and whether SNO and KamLAND results select, or do not select, the LMA solar 
neutrino solution. 
If $\nu_\mu \rightarrow \nu_e$ oscillations are observed, 
then $\sin^2 2\theta_{13} > 0.01$. If in addition the LMA solution correctly 
describes the solar neutrino deficit, then a high--performance neutrino 
factory hits the physics jackpot, addressing the pressing questions: 
Is there significant CP violation in the lepton sector? What is the sign of 
$\Delta m^2_{32}$? What is the precise value of $\sin^2 2\theta_{13}$?
In this scenario it might also be possible to make some progress with 
superbeams, but this requires further study. 

In the remaining scenarios (the LMA solution does not describe the 
solar neutrino deficit and/or $\sin^2 2\theta_{13} < 0.01$) progress 
on determining the mixing matrix elements and neutrino mass spectrum 
will be harder, but neutrino factories still offer the possibility 
of learning more, and may indeed offer the only way of probing very 
small values of $\sin^2 2\theta_{13}$.

\section{Non-Oscillation Physics}

Although neutrino oscillations provide the primary motivation for 
the development of a neutrino factory, we should not neglect the 
other neutrino physics that could be pursued at a very intense 
high energy neutrino source. A high--performance neutrino factory 
would produce beams a few hundred meters downstream of the storage ring 
that are a factor O($10^4$) more intense than existing 
conventional neutrino beams! This would have a tremendous impact 
on non--oscillation neutrino physics. For example, we can imagine 
the use of 
silicon pixel targets, or hydrogen and deuterium polarized targets, 
together with compact high--granularity detectors with good particle 
identification. Some examples of experiments that might be attractive 
at a neutrino factory have been discussed in Ref.~\cite{physics_study}:
\begin{itemize}
\item Precise measurements of the detailed structure of the nucleon 
for each parton flavor, including the changes that occur in a 
nuclear environment.
\item A first measurement of the nucleon spin structure with neutrinos.
\item Charm physics with several million tagged particles. Note that 
charm production becomes significant for neutrino factory energies 
above 20~GeV.
\item Precise measurements of Standard Model parameters: $\alpha_s$, 
$\sin^2\theta_W$, and the $V_{CKM}$ matrix elements.
\item Searches for exotic phenomena such as neutrino magnetic moments,
 anomalous couplings to the tau--lepton, and additional neutral leptons.
\end{itemize}
The physics opportunities at neutrino factories are clearly 
not limited to neutrino oscillations.

\section{Final remarks}

In the next few years the particle physics community must decide 
which neutrino physics facilities should be proposed for the 
era beyond the next generation of experiments. The recent evidence 
for neutrino oscillations vastly increases the motivation for a 
large scale endeavor. Neutrino factories offer the possibility of 
completely determining the mixing matrix and the pattern of neutrino 
masses, determining whether there is significant CP violation in the 
lepton sector, clarifying the oscillation framework, determining whether 
there are light sterile neutrinos, and consolidating 
or changing our understanding of solar neutrino oscillations.
We can hope, although not guarantee, that neutrino factory measurements 
will enable us to discriminate between GUTs, or point the way to 
alternative theories that lead us to an understanding of the origin of 
quark and lepton flavors.
Finally, since neutrino oscillations require physics 
beyond the Standard Model, there is the very real possibility that 
something unexpected will be discovered at a neutrino factory.

\bigskip
\subsection*{Acknowledgments}

I am indebted to all those who have contributed
to the extensive literature on physics at a neutrino 
factory, and in particular to the many participants of the 
recent 6 months physics study at Fermilab, and to Mike Shaevitz who 
requested and encouraged the study. 
I am particularly indebted to Vernon Barger, Rajendran Raja, and 
Kerry Whisnant, with whom I have collaborated on a sequence of calculations. 
Special thanks to Carl Albright for providing 
invaluable feedback on the manuscript. Finally, much credit goes to the 
Neutrino Factory and 
Muon Collider Collaboration without which there would be little prospect of 
developing muon sources with the intensity required for a neutrino factory. 
This work was supported at Fermilab under grant US DOE DE-AC02-76CH03000.

\clearpage



\begin{thebibliography}{99}
\bibitem{nufact_old}
Note that an older concept, acsribed to Kushkarev, Wojcicki, and 
Collins, was to inject pions of the desired energy into a storage 
ring with straight--sections, the pions decay in situ to create a 
captured muon beam. However, the resulting neutrino beam intensity 
is a factor of 1000 less than obtained with the modern neutrino factory 
scheme.
\end{thebibliography}
\end{document}